\documentclass[conference]{IEEEtran}
\usepackage{cite}

\ifCLASSINFOpdf
\usepackage[pdftex]{graphicx}
\else
\fi
\usepackage{amsmath}
\usepackage{amsfonts} 
\usepackage{amsthm}
\usepackage{amssymb}
\usepackage[boxruled]{algorithm2e}
\usepackage{amsthm} 
\usepackage{mathtools} 

\usepackage{BOONDOX-ds}

\usepackage{bbm}

\usepackage{kbordermatrix}

\newcommand{\pfun}{\mathop{\hbox{$\to$\kern-7pt\raise.9pt\hbox{\scalebox{1}[.55]{$|$}}\kern4pt} }}

\usepackage{array}

\begin{document}

\title{TabulaROSA: Tabular Operating System Architecture for Massively Parallel \\ Heterogeneous Compute Engines}

\author{\IEEEauthorblockN{Jeremy Kepner$^{1-4}$, Ron Brightwell$^5$, Alan Edelman$^{2,3}$, 
Vijay Gadepally$^{1,2,4}$, Hayden Jananthan$^{1,4,6}$, \\ Michael Jones$^{1,4}$, Sam Madden$^2$, Peter Michaleas$^{1,4}$, Hamed Okhravi$^4$, Kevin Pedretti$^5$, \\ Albert Reuther$^{1,4}$, Thomas Sterling$^7$, Mike Stonebraker$^2$
\\
\IEEEauthorblockA{$^1$MIT Lincoln Laboratory Supercomputing Center, $^2$MIT Computer Science \& AI Laboratory, \\ $^3$MIT Mathematics Department, $^4$MIT Lincoln Laboratory Cyber Security Division, \\
$^5$Sandia National Laboratories Center for Computational Research, \\ $^6$Vanderbilt University Mathematics Department, \\
$^7$Indiana University Center for Research in Extreme Scale Technologies
}}}
\maketitle

\begin{abstract}
The rise in computing hardware choices is driving a reevaluation of operating systems.  The traditional role of an operating system controlling the execution of its own hardware is evolving toward a model whereby the controlling processor is distinct from the compute engines that are performing most of the computations.  In this context, an operating system can be viewed as software that brokers and tracks the resources of the compute engines and is akin to a database management system.  To explore the idea of using a database in an operating system role, this work defines key operating system functions in terms of rigorous mathematical semantics (associative array algebra) that are directly translatable into database operations.  These operations possess a number of mathematical properties that are ideal for parallel operating systems by guaranteeing correctness over a wide range of parallel operations.  The resulting operating system equations provide a mathematical specification for a Tabular Operating System Architecture (TabulaROSA) that can be implemented on any platform. Simulations of forking in TabularROSA are performed using an associative array implementation and compared to Linux on a 32,000+ core supercomputer.  Using over 262,000 forkers managing over 68,000,000,000 processes, the simulations show that TabulaROSA has the potential to perform operating system functions on a massively parallel scale.  The TabulaROSA simulations show 20x higher performance as compared to Linux while managing 2000x more processes in fully searchable tables.
\end{abstract}

%
\IEEEpeerreviewmaketitle

\section{Introduction}
\let\thefootnote\relax\footnotetext{This material is based upon work supported by the Assistant Secretary of Defense for Research and Engineering under Air Force Contract No. FA8721-05-C-0002 and/or FA8702-15-D-0001. Any opinions, findings, conclusions or recommendations expressed in this material are those of the author(s) and do not necessarily reflect the views of the Assistant Secretary of Defense for Research and Engineering.}

\begin{figure}[ht]
\centering
\includegraphics[width=\columnwidth]{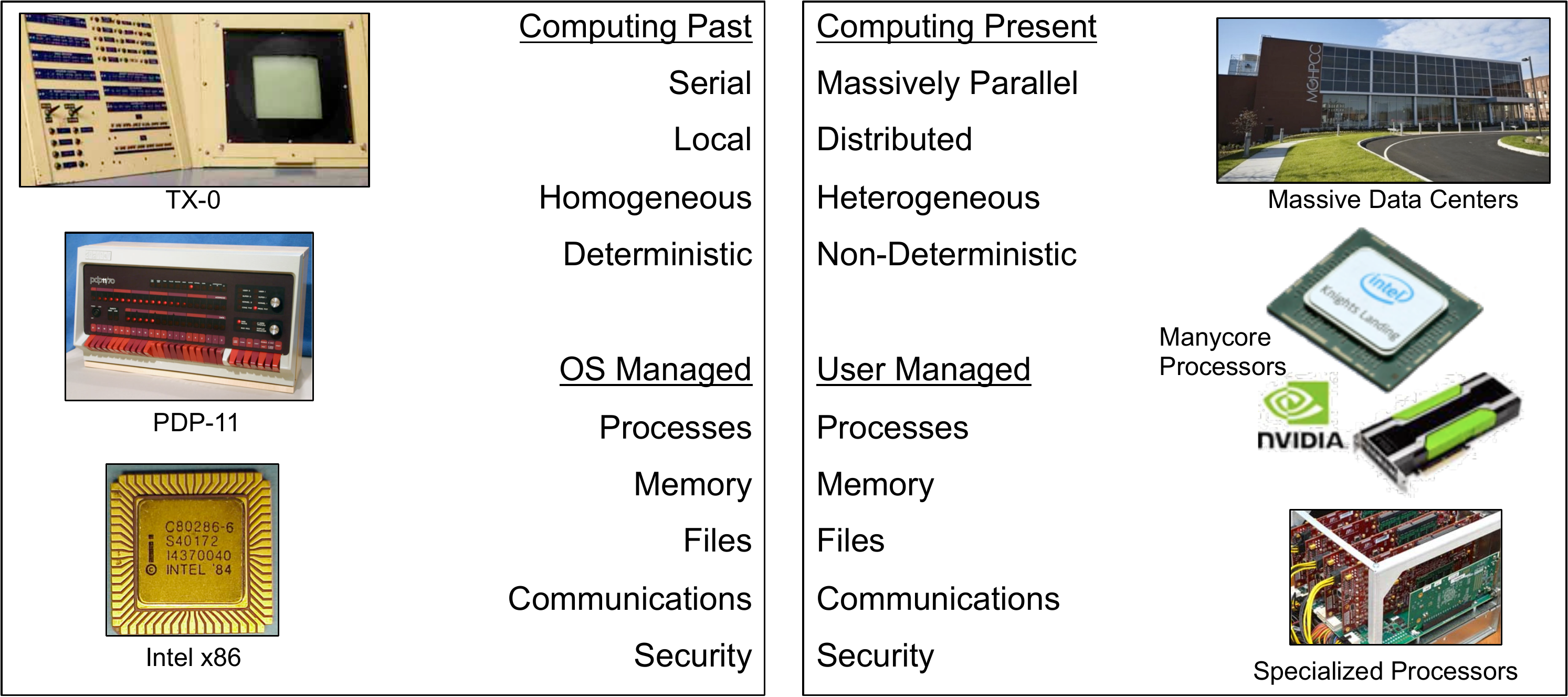}
\caption{Current operating systems can trace their lineage back to the first computers and still have many features that are designed for that era.  Modern computers are very different and currently require the user to perform most operating system functions.}
\label{fig:OShistory}
\end{figure}

Next generation computing hardware is increasingly purpose built for simulation \cite{sterling2017high}, data analysis  \cite{song2016novel}, and machine learning \cite{lecun2015deep}.  The rise in computing hardware choices: general purpose central processing units (CPUs), vector processors, graphics processing units (GPUs), tensor processing units (TPUs), field programmable gate arrays (FPGAs), optical computers, and quantum computers is driving a reevaluation of operating systems. Even within machine learning, there has been a trend toward developing more specialized processors for different stages and types of deep neural networks (e.g., training vs inference, dense vs sparse networks).  Such hardware is often massively parallel, distributed, heterogeneous, and non-deterministic, and must satisfy a wide range of security requirements. Current mainstream operating systems (OS) can trace their lineages back 50 years to computers designed for basic office functions running on serial, local, homogeneous, deterministic hardware operating in benign environments (see Figure~\ref{fig:OShistory}).  Increasingly, these traditional operating systems are bystanders at best and impediments at worse to using purpose-built processors.  This trend is illustrated by the current GPU programming model whereby the user engages with a conventional OS to acquire the privilege of accessing the GPU to then implement most OS functions (managing memory, processes, and IO) inside their own application code.  

  The role of an operating system controlling the execution of its own hardware is evolving toward a model whereby the controlling processor is distinct from the compute engines that are performing most of the computations \cite{beckman2012exascale,schwarzkopf2015operating,laplante2016rethinking}.
Traditional operating systems like Linux are built to execute a shared kernel on homogeneous cores  \cite{torvalds1997linux}.  Popcorn Linux \cite{barbalace2015popcorn} and K2 \cite{lin2014k2} run multiple Linux instances on heterogeneous cores. Barrelfish \cite{baumann2009multikernel} and FOS (Factored Operating System) \cite{wentzlaff2009factored} aim to support many heterogeneous cores over a distributed system.  NIX \cite{ballesteros2012nix}, based on Plan 9 \cite{pike1995plan}, relaxes the requirement on executing a kernel on every core by introducing application cores.  Helios \cite{nightingale2009helios}, a derivative from Singularity \cite{fahndrich2006language}, reduces the requirements one step further by using software isolation instead of address space protection. Thereby, neither a memory management unit nor a privileged mode is required.  In order to address the increasing role of accelerators, the M3 operating system goes further and removes all requirements on processor features \cite{asmussen2016m3}.

In this context, an OS can be viewed as software that brokers and tracks the resources of compute engines. Traditional supercomputing schedules currently fill the role of managing heterogeneous resources but have inherent scalability limitations \cite{reuther2018scalable}.  In many respects, this new operating system role is akin to the traditional role of a database management system (DBMS) and suggests that databases may be well suited to operating system tasks for future hardware architectures.  To explore this hypothesis, this work defines key operating system functions in terms of rigorous mathematical semantics (associative array algebra) that are directly translatable into database operations.  Because the mathematics of database table operations are based on a linear system over the union and intersection semiring, these operations possess a number of mathematical properties that are ideal for parallel operating systems by guaranteeing correctness over a wide range of parallel operations.  The resulting operating system equations provide a mathematical specification for a Tabular Operating System Architecture (TabulaROSA) that can be implemented on any platform. Simulations of selected TabularROSA functions are performed with an associative array implementation on state-of-the-art, highly parallel processors.  The measurements show that TabulaROSA has the potential to perform operating system functions on a massively parallel scale with 20x higher performance.

\section{Standard OS and DBMS Operations}

\begin{figure}[ht]
\centering
\includegraphics[width=\columnwidth]{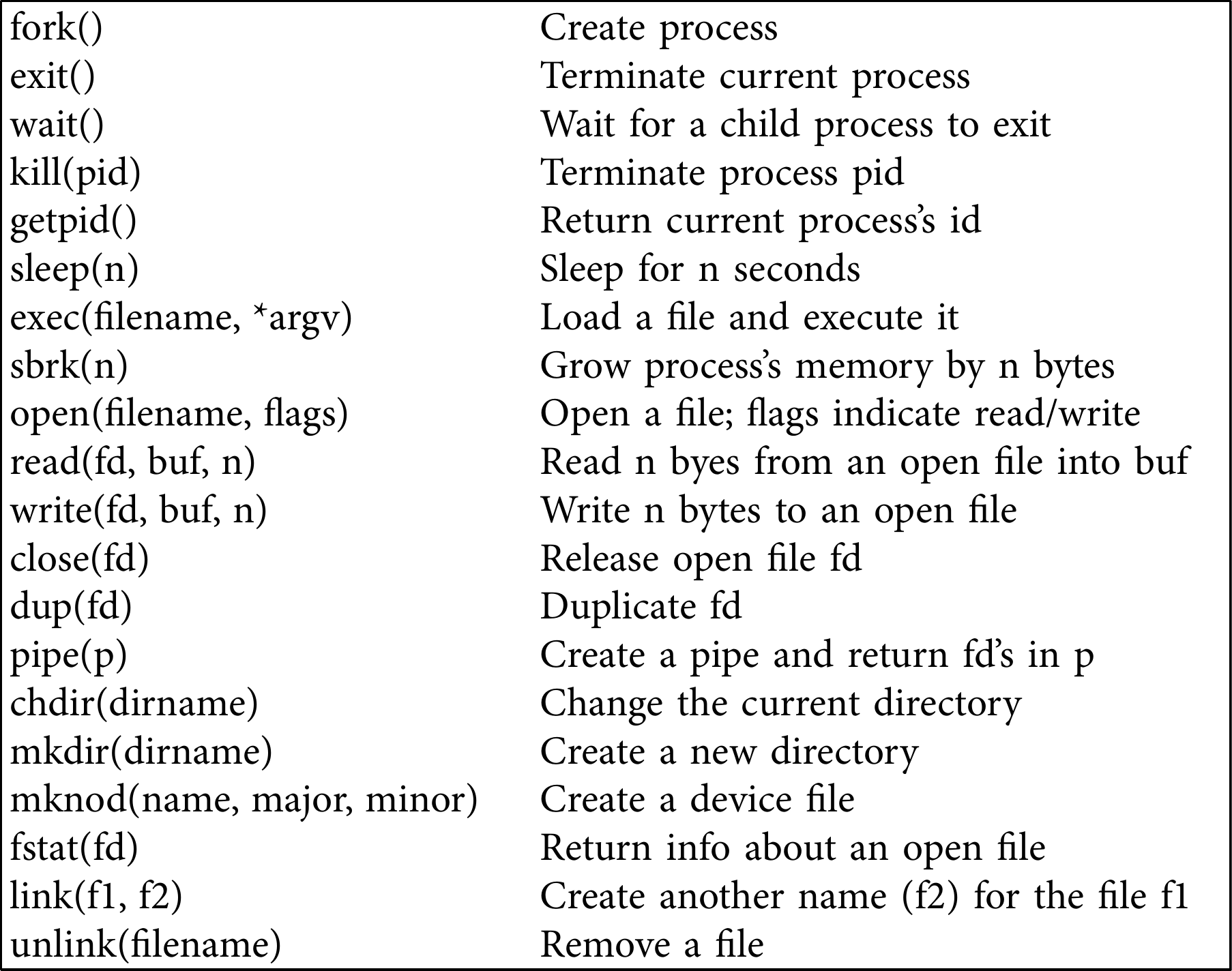}
\caption{(Adapted from \cite{cox2011xv6}). Xv6 operating system kernel functions.}
\label{fig:xv6}
\end{figure}

  TabulaROSA seeks to explore the potential benefits of implementing OS functions in way that leverages the power and mathematical properties of database systems.  This exploration begins with a brief description of standard OS and DBMS functions. Many concepts in modern operating systems can trace their roots to the very first time-sharing computers. Unix was first built in 1970 on the Digital Equipment Corporation (DEC) Programmed Data Processor (PDP)\cite{dennis1964multiuser,bell1970new}, which was based on the MIT Lincoln Laboratory Transistorized eXperimental computer zero (TX-0)\cite{mitchell1956tx,clark1957lincoln,mccarthy1963time,myer1968design}.   Modern operating systems, such as Linux, are vast, but their core concepts can be reduced to a manageable set of operations, such as those captured in the Xv6 operating system \cite{cox2011xv6}.  Xv6 is a teaching operating system developed in 2006 for MIT's operating systems course 6.828: Operating System Engineering.  Xv6 draws inspiration from  Unix V6 \cite{ritchie1978unix} and  Lions' Commentary on UNIX, 6th Edition \cite{lions2000lions}. An elegant aspect of Unix is that its primary interface is the C language, which is also the primary language that programmers use to develop applications in Unix.  Xv6 can be summarized by its C language kernel system calls that define its interface to the user programmer (see Figure~\ref{fig:xv6}). This subset of system calls is representative of the services that are core to many Unix based operating systems and serves as a point of departure for TabulaROSA.  At a deeper level, many of the Xv6 operating system functions can be viewed as adding, updating, and removing records from a series of C data structures that are similar to purpose-built database tables. 

\begin{figure}[ht]
\centering
\includegraphics[width=\columnwidth]{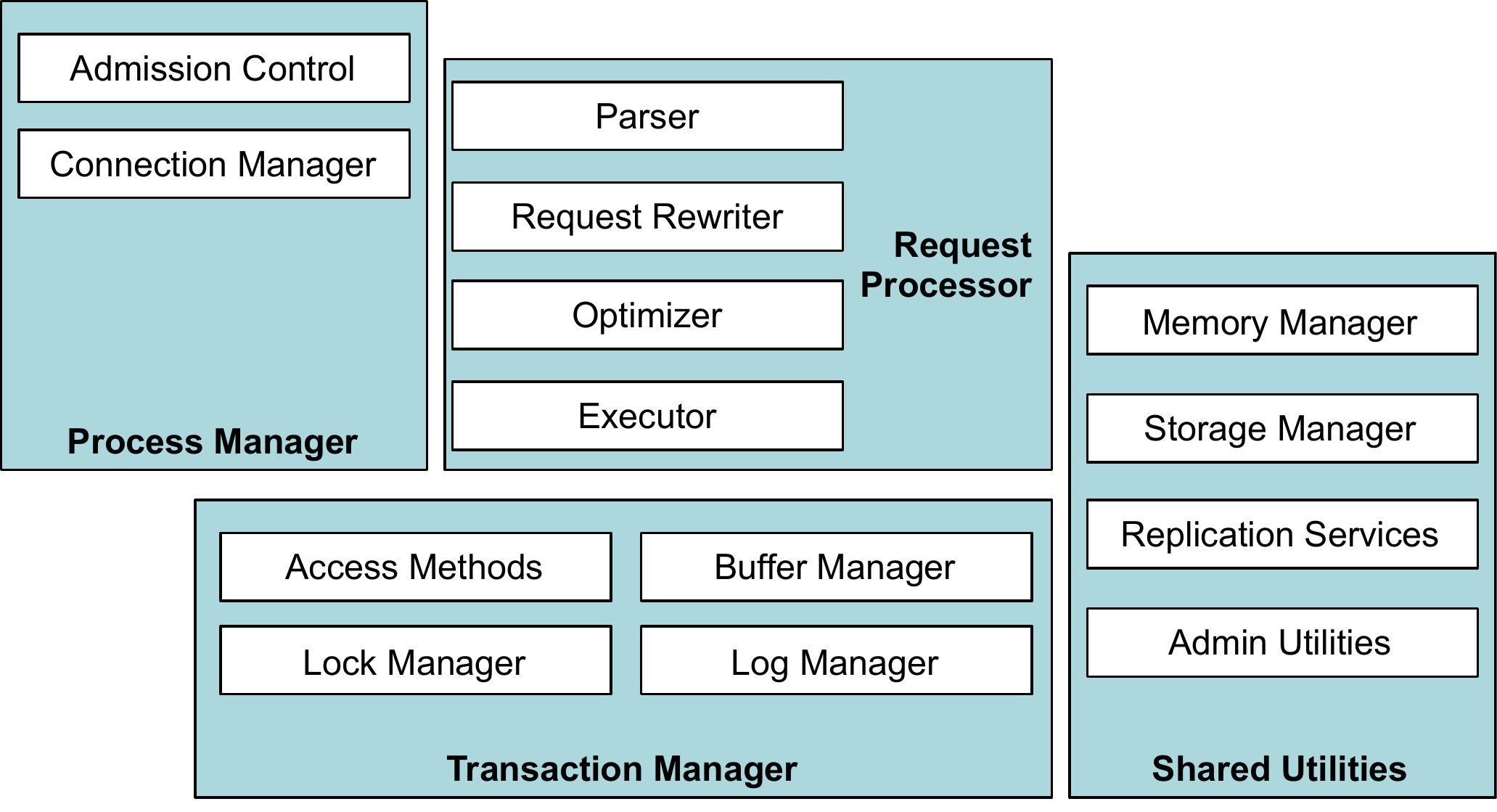}
\caption{(Adapted from \cite{hellerstein2005readings}). Database management systems provide many of the functions found in an operating system. Nearly all functions are managed via manipulations of database tables.}
\label{fig:DBMS}
\end{figure}

Modern database systems are designed to perform many of the same functions of purpose-built data analysis, machine learning, and simulation hardware.  The key tasks of many modern data processing systems can be summarized as follows \cite{stonebrakerseven}
\begin{itemize}
\item Ingesting data from operational data systems
\item Data cleaning
\item Transformations
\item Schema integration
\item Entity consolidation
\item Complex analytics
\item Exporting unified data to downstream systems
\end{itemize}
To meet these requirements, database management systems perform many operating system functions (see Figure~\ref{fig:DBMS}) \cite{hellerstein2005readings}.  

An elegant aspect of many DBMS is that their primary data structures are tables that are also the primary data structures programmers use to develop applications on these systems.  In many databases, these table operations can be mapped onto well-defined mathematical operations with known mathematical properties.  For example, relational (or SQL) databases \cite{Stonebraker1976,date1989guide,elmasri2010fundamentals} are described by relational algebra \cite{codd1970relational,maier1983theory,Abiteboul1995} that corresponds to the union-intersection semiring ${\cup}.{\cap}$ \cite{jananthan2017polystore}.  Triple-store databases (NoSQL) \cite{DeCandia2007,LakshmanMalik2010,George2011,Wall2015} and analytic databases (NewSQL) \cite{Stonebraker2005,Kallman2008,Balazinska2009,StonebrakerWeisberg2013,Hutchison2015,gadepally2015graphulo}  follow similar mathematics \cite{kepner2016associative}.  The table operations of these databases are further encompassed by associative array algebra, which brings the beneficial properties of matrix mathematics and sparse linear systems theory, such as closure, commutativity, associativity, and distributivity \cite{kepnerjananthan}.

The aforementioned mathematical properties provide strong correctness guarantees that are independent of scale and particularly helpful when trying to reason about massively parallel systems. Intersection $\cap$ distributing over union $\cup$ is essential to database query planning and parallel query execution over partioned/sharded database tables \cite{booth1976distributed,shaw1980relational,stonebraker1986case,barroso2003web,curino2010schism,pavlo2012skew,corbett2013spanner}.  Similarly, matrix multiplication distributing over matrix addition ensures the correctness of massively parallel implementations on the world's largest supercomputers \cite{dongarra2003linpack} and machine learning systems \cite{moller1993scaled,werbos1994roots,chetlur2014cudnn}.  In software engineering, the scalable commutativity rule guarantees the existence of a conflict-free (parallel) implementation \cite{clements2015scalable,clements2017scalable,bhat2017designing}.

\section{Associative Array Algebra}

  The full mathematics of associative arrays and the ways they encompass matrix mathematics and relational algebra are described in the aforementioned references \cite{jananthan2017polystore,kepner2016associative,kepnerjananthan}.  Only the essential mathematical properties of associative arrays necessary for describing TabulaROSA are reviewed here.  The essence of associative array algebra is three operations: element-wise addition (database table union), element-wise multiplication (database table intersection), and array multiplication (database table transformation).  In brief, an associative array $\mathbf{A}$ is defined as a mapping from sets of keys to values
$$
  \mathbf{A}: K_1 \times K_2 \to \mathbb{V}
$$
where $K_1$ are the row keys and $K_2$ are the column keys and can be any sortable set, such as integers, real numbers, and strings. The row keys are equivalent to the sequence ID in a relational database table or the process ID of file ID in an OS data structure.  The column keys are equivalent to the column names in a database table and the field names in an OS data structure.  $\mathbb{V}$ is a set of values that forms a semiring $(\mathbb{V},\oplus,\otimes,0,1)$ with addition operation $\oplus$, multiplication operation $\otimes$, additive identity/multiplicative annihilator 0, and multiplicative identity 1. The values can take on many forms, such as numbers, strings, and sets. One of the most powerful features of associative arrays is that addition and multiplication can be a wide variety of operations.  Some of the common combinations of addition and multiplication operations that have proven valuable are standard arithmetic addition and multiplication ${+}.{\times}$, the aforementioned union and intersection ${\cup}.{\cap}$, and various tropical algebras that are important in finance \cite{klemperer2010product,baldwin2016understanding,masontropical} and neural networks \cite{Kepner2017graphblasDNN}: ${\max}.{+}$, ${\min}.{+}$, ${\max}.{\times}$, ${\min}.{\times}$, ${\max}.{\min}$, and ${\min}.{\max}$.

The construction of an associative array is denoted
$$
  \mathbf{A} = \mathbb{A}(\mathbf{k}_1,\mathbf{k}_2,\mathbf{v})
$$
where $\mathbf{k}_1$, $\mathbf{k}_2$, and $\mathbf{v}$ are vectors of the row keys, column keys, and values of the nonzero elements of $\mathbf{A}$.  When the values are 1 and there is only one nonzero entry per row or column, this associative array is denoted
$$
  \mathbb{I}(\mathbf{k}_1,\mathbf{k}_2) = \mathbb{A}(\mathbf{k}_1,\mathbf{k}_2,1)
$$
and when $\mathbb{I}(\mathbf{k}) = \mathbb{I}(\mathbf{k},\mathbf{k})$, this is array is referred to as the identity.

Given associative arrays $\mathbf{A}$, $\mathbf{B}$, and $\mathbf{C}$, element-wise addition is denoted
$$
   \mathbf{C} = \mathbf{A} \oplus \mathbf{B}
$$
or more specifically
$$
   \mathbf{C}(k_1,k_2) = \mathbf{A}(k_1,k_2) \oplus \mathbf{B}(k_1,k_2)
$$
where $k_1 \in K_1$ and $k_2 \in K_2$. Similarly, element-wise multiplication is denoted
$$
   \mathbf{C} = \mathbf{A} \otimes \mathbf{B}
$$
or more specifically
$$
   \mathbf{C}(k_1,k_2) = \mathbf{A}(k_1,k_2) \otimes \mathbf{B}(k_1,k_2)
$$
Array multiplication combines addition and multiplication and is written
$$
   \mathbf{C} = \mathbf{A} \mathbf{B} = \mathbf{A} {\oplus}.{\otimes} \mathbf{B}
$$
or more specifically
$$
   \mathbf{C}(k_1,k_2) = \bigoplus_k \mathbf{A}(k_1,k) \otimes \mathbf{B}(k,k_2)
$$
where $k$ corresponds to the column key of $\mathbf{A}$ and the row key of $\mathbf{B}$. Finally, the array transpose is denoted
$$
   \mathbf{A}(k_2,k_1) = \mathbf{A}^{\sf T}(k_1,k_2)
$$
The above operations have been found to enable a wide range of database algorithms and matrix mathematics while also preserving several valuable mathematical properties that ensure the correctness of parallel execution.  These properties include commutativity
\begin{eqnarray*}
   \mathbf{A} \oplus \mathbf{B} &=& \mathbf{B} \oplus \mathbf{A} \\
   \mathbf{A} \otimes \mathbf{B} &=& \mathbf{B} \otimes \mathbf{A} \\
   (\mathbf{A} \mathbf{B})^{\sf T} &=& \mathbf{B}^{\sf T} \mathbf{A}^{\sf T}
\end{eqnarray*}
associativity
\begin{eqnarray*}
   (\mathbf{A} \oplus \mathbf{B}) \oplus \mathbf{C} &=& \mathbf{A} \oplus (\mathbf{B} \oplus \mathbf{C}) \\
   (\mathbf{A} \otimes \mathbf{B}) \otimes \mathbf{C} &=& \mathbf{A} \otimes (\mathbf{B} \otimes \mathbf{C}) \\
   (\mathbf{A}  \mathbf{B})  \mathbf{C} &=& \mathbf{A}  (\mathbf{B}  \mathbf{C})
\end{eqnarray*}
distributivity
\begin{eqnarray*}
   \mathbf{A} \otimes (\mathbf{B} \oplus \mathbf{C}) &=& (\mathbf{A} \otimes \mathbf{B}) \oplus (\mathbf{A} \otimes \mathbf{C}) \\
   \mathbf{A} (\mathbf{B} \oplus \mathbf{C}) &=& (\mathbf{A}  \mathbf{B}) \oplus (\mathbf{A}  \mathbf{C}) 
\end{eqnarray*}
and the additive and multiplicative identities
$$
   \mathbf{A} \oplus \mathbs{0} = \mathbf{A} ~~~~~~~~~~~~ \mathbf{A} \otimes \mathbs{1} = \mathbf{A} ~~~~~~~~~~~~ \mathbf{A} \mathbb{I} = \mathbf{A}
$$
where $\mathbs{0}$ is an array of all 0, $\mathbs{1}$ is an array of all 1, and $\mathbb{I}$ is an array with 1 along its diagonal.  Furthermore, these arrays possess a multiplicative annihilator
$$
   \mathbf{A} \otimes \mathbs{0} = \mathbs{0} ~~~~~~~~~~~~ \mathbf{A} \mathbs{0} = \mathbs{0}
$$
Most significantly, the properties of associative arrays are determined by the properties of the value set $\mathbb{V}$.  In other words, if $\mathbb{V}$ is linear (distributive), then so are the corresponding associative arrays.

\section{TabulaROSA Mathematics}

\begin{figure}[ht]
\centering
\includegraphics[width=0.85\columnwidth]{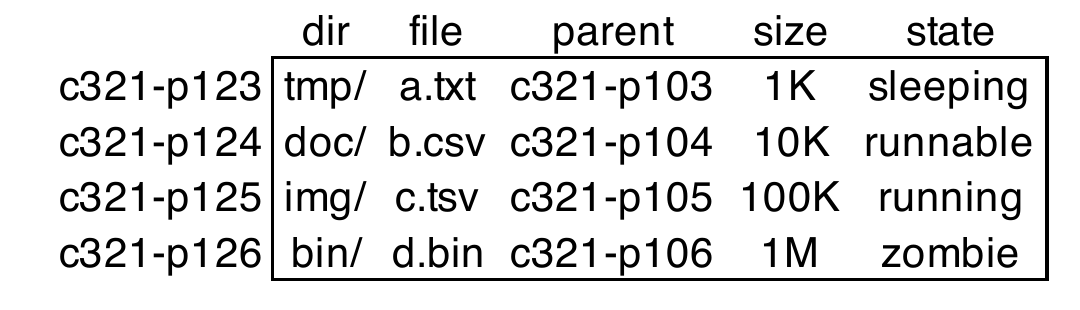}
\includegraphics[width=\columnwidth]{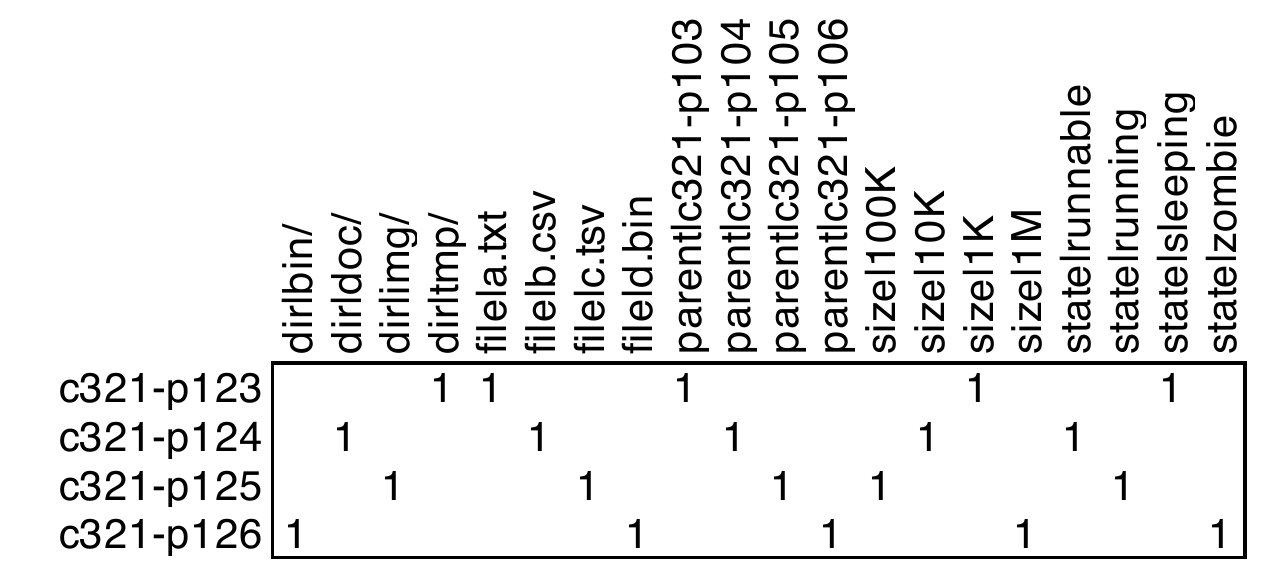}
\caption{Notional examples of dense (top) and sparse (bottom) schemas for the distributed global process associative array $\mathbf{P}$, where the rows are the process IDs and the columns are metadata describing each process.}
\label{fig:Pschema}
\end{figure}

  There are many possible ways of describing the Xv6 OS functions in terms of associative arrays.  One possible approach begins with the following definitions.   $\mathbf{P}$ is the distributed global process associative array, where the rows are the process IDs and the columns are metadata describing each process.  In Xv6, there are approximately ten metadata fields attributed to each process.  Notional examples of dense and sparse schemas for $\mathbf{P}$ are shown in Figure~\ref{fig:Pschema}.  In this analysis, a hybrid schema is assumed as it naturally provides fast search on any row or column, enables most OS operations to be performed with array multiplication, and allows direct computation on numeric values.   $\mathbf{p}$ is a vector containing one or more unique process IDs and is implicitly the output of the getpid() accessor function or all processes associated with the current context.  Similarly, $\dot{\mathbf{p}}$ is implicitly the output of allocproc().

Associative array specifications of all the Xv6 functions listed in Figure~\ref{fig:xv6} are provided in Appendix A.  Perhaps the most important of these functions is fork(), which is used to create new processes and is described mathematically as follows

\noindent --------------------------------------------------------------------

\noindent $\dot{\mathbf{p}}$ = fork() ~~~~~~ \# Function for creating processes

\noindent -- -- -- -- -- -- -- -- -- -- -- -- -- -- -- -- -- -- -- -- -- -- -- -- -- -- -- 

$\dot{\mathbf{p}} ~~ = $ allocproc() ~~~~~~~~\# Create new process IDs

$\dot{\mathbf{P}} ~~ = ~ \mathbb{I}(\dot{\mathbf{p}},\mathbf{p}) \mathbf{P}$ ~~~~~~~~~\# Create new $\dot{\mathbf{P}}$ from $\mathbf{P}$

$\dot{\mathbf{P}} ~ {\oplus}{=} ~ \mathbb{I}(\dot{\mathbf{p}},{\rm parent}|\mathbf{p})$ ~~ \# Add parent identifiers

$\dot{\mathbf{P}} ~ {\oplus}{=} ~ \mathbb{I}(\mathbf{p},{\rm child}|\dot{\mathbf{p}})$ ~~~~ \# Add child identifiers

$\mathbf{P} ~ {\oplus}{=} ~ \dot{\mathbf{P}}$ ~~~~~ \# Add new processes to global table

\noindent --------------------------------------------------------------------

\noindent where $|$ implies concatenation with a separator such as $|$.  The above mathematical description of fork() can also be algebraically compressed into the following single equation
$$
\mathbf{P} ~ {\oplus}{=} ~ \mathbb{I}(\dot{\mathbf{p}},\mathbf{p}) \mathbf{P} \oplus \mathbb{I}(\dot{\mathbf{p}},{\rm parent}|\mathbf{p}) \oplus \mathbb{I}(\mathbf{p},{\rm child}|\dot{\mathbf{p}})
$$
Additionally, forking the same process into multiple new processes can be done by adding nonzero rows to $\mathbb{I}(\dot{\mathbf{p}})$.  Likewise, combining existing processes into a single new process can be done by adding nonzero entries to any row of $\mathbb{I}(\dot{\mathbf{p}},\mathbf{p})$.  Thus, the associative array representation of fork() can accommodate a range of fork() operations that normally require distinct implementations.  The above equation is independent of scale and describes a method for simultaneously forking many processes at once.  The above equation is linear,  so many of the operations can be ordered in a variety of ways while preserving correctness.  Because dominant computation is the array multiplication  $\mathbb{I}(\dot{\mathbf{p}},\mathbf{p}) \mathbf{P}$, the performance and scalability of forking in TabulaROSA can be estimated by simulating this operation.

\section{Simulation Results}

  The associative array representation of fork() indicates that the core computational driver in a massively parallel implementation would be associative array multiplication  $\mathbb{I}(\dot{\mathbf{p}},\mathbf{p}) \mathbf{P}$.  This operation can be readily simulated with the D4M (Dynamic Distributed Dimensional Data Model) implementation of associative arrays that are available in a variety of programming languages (d4m.mit.edu)\cite{Kepner2012,chen2016julia,kepnerjananthan}.  In the simulation, each forker constructs a $2^{18} \times 2^{18}$ sparse associative array $\mathbf{P}$ with approximately 10 nonzero entries per row.  The row keys of $\mathbf{P}$ are the globally unique process IDs $\mathbf{p}$.  The column keys of $\mathbf{P}$ are strings that are surrogates for process ID metadata fields.  The fork() operation is simulated by array multiplication of $\mathbf{P}$ by a $2^{18} \times 2^{18}$ permutation array $\mathbb{I}(\dot{\mathbf{p}},\mathbf{p})$.  Each row and column in $\mathbb{I}(\dot{\mathbf{p}},\mathbf{p})$ has one randomly assigned nonzero entry that corresponds to the mapping of the current process IDs in $\mathbf{p}$ to the new process IDs in $\dot{\mathbf{p}}$.  The full sparse representation is the most computationally challenging as it represents the case whereby all unique metadata have their own column and are directly searchable. For comparison, the Linux fork() command is also timed, and it is assumed that the maximum number of simultaneous processes on a compute node is the standard value of $2^{16}$.

\begin{figure}[ht]
\centering
\includegraphics[width=\columnwidth]{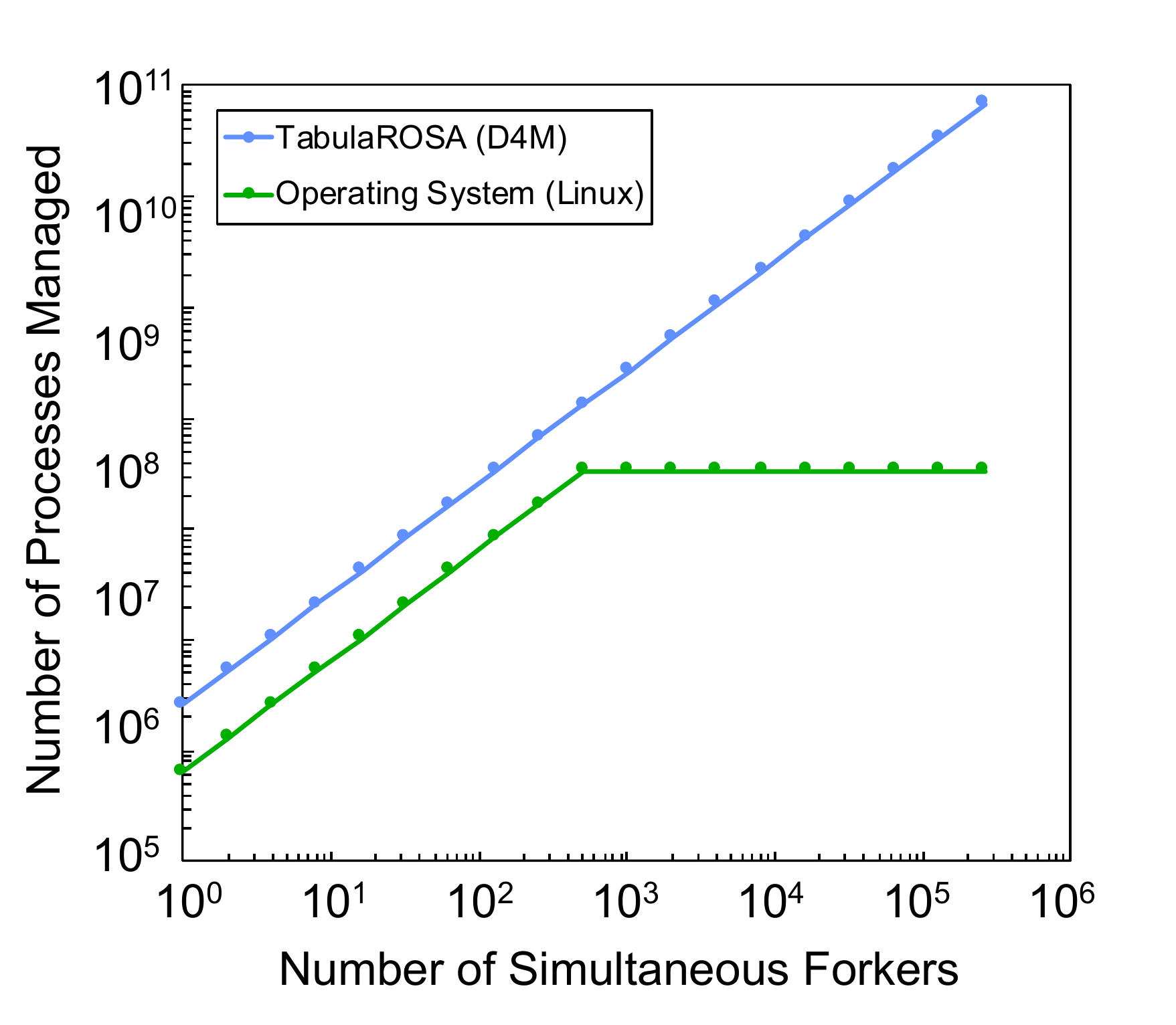}
\caption{Number of processes managed vs. total number of forkers for the TabulaROSA D4M simulation and the Linux operating system running on a 32,000+ core system.}
\label{fig:ProcessesManaged}
\end{figure}

\begin{figure}[ht]
\centering
\includegraphics[width=\columnwidth]{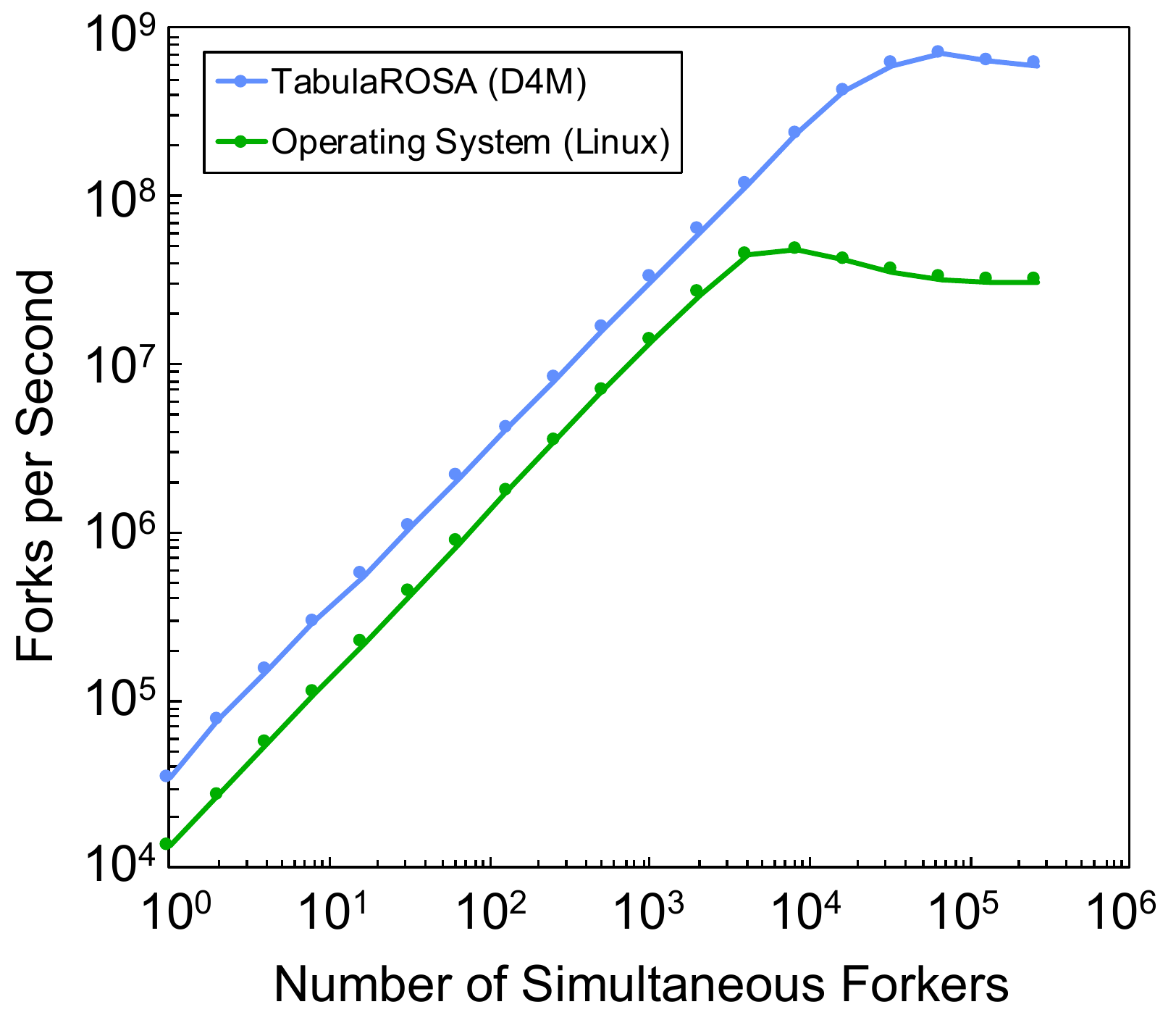}
\caption{Fork rate vs. total number of forkers for the TabulaROSA D4M simulation and the Linux operating system running on a 32,000+ core system.}
\label{fig:ForkRate}
\end{figure}

  Both the D4M simulation and the Linux fork() comparison were run in parallel on a supercomputer consisting of 648 compute nodes, each with at least 64 Xeon processing cores, for a total of 41,472 processing cores.  The results are shown in Figures~\ref{fig:ProcessesManaged} and \ref{fig:ForkRate}.  In both cases, a single forker was run on 1, 2, 4,\ldots, and 512 compute nodes, followed by running 2, 4, \ldots, and 512 forkers  on each of the 512 compute nodes to achieve a maximum of 262,144 simultaneous forkers.  This pleasingly parallel calculation would be expected to scale linearly on a supercomputer.
  
  The number of processes managed in the D4M simulation grows linearly with the number of forkers, while in the Linux fork() comparison the number of processes managed grows linearly with the number of compute nodes. Figure~\ref{fig:ProcessesManaged} shows the total number of processes managed, which for D4M is $2^{18}$ times the number of forkers with a largest value of $2^{36}$ or over 68,000,000,000. For the Linux fork() comparison, the total number of processes managed is $2^{16}$ times the number of nodes with a largest value of $2^{25}$.  In both computations, the number of processes managed could be increased. For these calculations, typical values are used.  Since associative array operations are readily translatable to databases using disk storage, the capacity of this approach is very large even when there are high update rates \cite{kepner2014achieving}.  
  
  Figure~\ref{fig:ForkRate} shows the rate at which processes forked in the D4M simulation grows linearly with the number of forkers and peaks at 64 forkers per node, which corresponds to the number of physical cores per node. In the Linux fork() comparison, the rate grows linearly with the number of forkers and peaks at 8 forkers per node. The largest fork rate for D4M in this simulation is approximately 800,000,000 forks per second.  The largest fork rate for the Linux fork() comparison is approximately 40,000,000 forks per second.

  The purpose of these comparisons is not to highlight any particular attribute of Linux, but merely to set the context for the D4M simulations, which demonstrate the size and speed potential of TabulaROSA.  
 
\section{Conclusion}

The rise in computing hardware choices: general purpose central processing units, vector processors, graphics processing units, tensor processing units, field programmable gate arrays, optical computers, and quantum computers are driving a reevaluation of operating systems.  The traditional role of an operating system controlling the execution on its own hardware is evolving toward a model whereby the controlling processor is completely distinct from the compute engines that are performing most of the computations.   In many respects, this new operating system role is akin to the traditional role of a database management system  and suggests that databases may be well suited to operating system tasks for future hardware architectures.  To explore this hypothesis, this work defines key operating system functions in terms of rigorous mathematical semantics (associative array algebra) that are directly translatable into database operations.  Because the mathematics of database table operations are based on a linear system over the union and intersection semiring, these operations possess a number of mathematical properties that are ideal for parallel operating systems by guaranteeing correctness over a wide range of parallel operations.  The resulting operating system equations provide a mathematical specification for a Tabular Operating System Architecture (TabulaROSA) that can be implemented on any platform.  Simulations of forking in TabularROSA are performed by using an associative array implementation and are compared to Linux on a 32,000+ core supercomputer.  Using over 262,000 forkers managing over 68,000,000,000 processes, the simulations show that TabulaROSA has the potential to perform operating system functions on a massively parallel scale.  The TabulaROSA simulations show 20x higher performance  compared to Linux, while managing 2000x more processes in fully searchable tables.


\section*{Acknowledgments}
%
%

The authors wish to acknowledge the following individuals for their contributions and support:
Bob Bond, Paul Burkhardt, Sterling Foster, Charles Leiserson, Dave Martinez, Steve Pritchard, Victor Roytburd, and Michael Wright.



\bibliographystyle{ieeetr}

\bibliography{aarabib}
%

\section*{Appendix A: TabulaROSA Specification}

$\mathbf{P}$ is the distributed global process associative array, where the rows are the process IDs and the columns are metadata describing each process.  In Xv6, approximately 10 metadata fields are attributed to each process.  Notional examples of dense and sparse schemas for $\mathbf{P}$ are shown in Figure~\ref{fig:Pschema}.  In this analysis, a hybrid schema is assumed as it naturally provides fast search on any row or column and allows most OS operations to be performed with array multiplication while still being able to perform direct computation numeric values.   $\mathbf{p}$ is a vector containing one or more unique process IDs and is implicitly the output of the getpid() accessor function or all processes associated with the current context.  Similarly, $\dot{\mathbf{p}}$ is implicitly the output of allocproc().  $\mathbf{F}$ is the distributed global files associative array where the rows are the file IDs and the columns are metadata describing each file and arguments for corresponding file operations.  $\mathbf{f}$ is a vector containing one or more unique file IDs, and $\mathbf{A_f}$ are the associative arrays corresponding to the contents in file identifiers $\mathbf{f}$.

\noindent --------------------------------------------------------------------

\noindent exec($\dot{\mathbf{F}}$) ~~~~~~~~~ \# Load files and execute them

\noindent -- -- -- -- -- -- -- -- -- -- -- -- -- -- -- -- -- -- -- -- -- -- -- -- -- -- -- 

$\dot{\mathbf{f}} = $ open($\dot{\mathbf{F}}$) ~~~~~~~~~~~~~ \# Open files

$\mathbf{P} ~ {\oplus}{=} ~ \mathbb{I}(\mathbf{p},\dot{\mathbf{f}}) \dot{\mathbf{F}}$ ~~~~~~~~~ \# Replace current instructions

\noindent --------------------------------------------------------------------

\noindent --------------------------------------------------------------------

\noindent $\mathbf{A}_{\rm buf}$ = read($\mathbf{A_{f}}$, row, col) ~ \# Read selected data

\noindent -- -- -- -- -- -- -- -- -- -- -- -- -- -- -- -- -- -- -- -- -- -- -- -- -- -- -- 

$\dot{\mathbf{f}} = $ open($\dot{\mathbf{F}}$) ~~~~~~~~~~~~~~ \# Open files

$\mathbf{A}_{\rm buf} = \mathbb{I}({\rm row}) \mathbf{A_{f}} \mathbb{I}({\rm col})$ ~ \# Select and copy to buffer

\noindent --------------------------------------------------------------------

\noindent --------------------------------------------------------------------

\noindent $\mathbf{A_{f}}$ = write($\mathbf{A}_{\rm buf}$, row, col) ~ \# Write selected data

\noindent -- -- -- -- -- -- -- -- -- -- -- -- -- -- -- -- -- -- -- -- -- -- -- -- -- -- -- 

$\dot{\mathbf{f}} = $ open($\dot{\mathbf{F}}$) ~~~~~~~~~~~~~~ \# Open files

$\mathbf{A_{f}} = \mathbb{I}({\rm row}) \mathbf{A}_{\rm buf} \mathbb{I}({\rm col})$ ~ \# Select and copy to file

\noindent --------------------------------------------------------------------

\noindent --------------------------------------------------------------------

\noindent $\dot{\mathbf{p}}$ = fork() ~~~~~~ \# Function for creating processes

\noindent -- -- -- -- -- -- -- -- -- -- -- -- -- -- -- -- -- -- -- -- -- -- -- -- -- -- -- 

$\dot{\mathbf{p}} ~~ = $ allocproc() ~~~~~~~~\# Create new process IDs

$\dot{\mathbf{P}} ~~ = ~ \mathbb{I}(\dot{\mathbf{p}},\mathbf{p}) \mathbf{P}$ ~~~~~~~~~\# Create new $\dot{\mathbf{P}}$ from $\mathbf{P}$

$\dot{\mathbf{P}} ~ {\oplus}{=} ~ \mathbb{I}(\dot{\mathbf{p}},{\rm parent}|\mathbf{p})$ ~~ \# Add parent identifiers

$\dot{\mathbf{P}} ~ {\oplus}{=} ~ \mathbb{I}(\mathbf{p},{\rm child}|\dot{\mathbf{p}})$ ~~~~ \# Add child identifiers

$\mathbf{P} ~ {\oplus}{=} ~ \dot{\mathbf{P}}$ ~~~~~ \# Add new processes to global table

\noindent --------------------------------------------------------------------

\noindent --------------------------------------------------------------------

\noindent exit() ~~~~~~~~~~~~ \# Exit current processes

\noindent -- -- -- -- -- -- -- -- -- -- -- -- -- -- -- -- -- -- -- -- -- -- -- -- -- -- -- 

$\mathbf{P} ~ {\oplus}{=} ~ \text{-}(\mathbb{I}(\mathbf{p}) \mathbf{P})$ ~~~~~~~~ \# Remove exiting processes

\noindent --------------------------------------------------------------------

\noindent --------------------------------------------------------------------

\noindent wait() ~~~~~~~~~~~~~ \# Wait for child processes to exit

\noindent -- -- -- -- -- -- -- -- -- -- -- -- -- -- -- -- -- -- -- -- -- -- -- -- -- -- -- 

$\dot{\mathbf{P}} = \mathbb{I}(\mathbf{p}) ~ \mathbf{P}  ~ \mathbb{I}({\rm child}|*) $ ~ \# Get child processes

while($\mathbf{P}$)

~~ $\dot{\mathbf{P}} = \mathbb{I}(\mathbf{p}) ~ \mathbf{P}  ~ \mathbb{I}({\rm child}|*) $ ~ \# Get exiting processes

\noindent --------------------------------------------------------------------

\noindent --------------------------------------------------------------------

\noindent kill($\mathbf{p}$) ~~~~~~~~~~~~~ \# Terminate processes

\noindent -- -- -- -- -- -- -- -- -- -- -- -- -- -- -- -- -- -- -- -- -- -- -- -- -- -- -- 

$\mathbf{P} ~ {\oplus}{=} ~ \text{-}(\mathbb{I}(\mathbf{p}) \mathbf{P})$ ~~~~~~ \# Remove processes to kill

\noindent --------------------------------------------------------------------

\noindent --------------------------------------------------------------------

\noindent $\mathbf{p}$ = getpid() ~~~~~~ \# Return current process IDs

\noindent -- -- -- -- -- -- -- -- -- -- -- -- -- -- -- -- -- -- -- -- -- -- -- -- -- -- -- 

$\mathbf{p} = {\rm row}(\mathbf{P}  ~ \mathbb{I}({\rm current}|*))$ ~ \# Get process IDs

\noindent --------------------------------------------------------------------
 
\noindent --------------------------------------------------------------------

\noindent $\dot{\mathbf{p}}$ = allocproc() ~~~ \# Return vector of new process IDs

\noindent -- -- -- -- -- -- -- -- -- -- -- -- -- -- -- -- -- -- -- -- -- -- -- -- -- -- -- 

$\dot{\mathbf{p}} = $ rand(), hash(), \ldots  ~ \# Create new process IDs

\noindent --------------------------------------------------------------------

\noindent --------------------------------------------------------------------

\noindent sleep($n$) ~~~~~~~~~~~ \# Sleep for $n$ seconds

\noindent -- -- -- -- -- -- -- -- -- -- -- -- -- -- -- -- -- -- -- -- -- -- -- -- -- -- -- 

$\mathbf{P} ~ {\oplus}{=} ~ \mathbb{A}(\mathbf{p},{\rm sleep},n)$ ~ \# Add $n$ seconds of sleep

\noindent --------------------------------------------------------------------

\noindent --------------------------------------------------------------------

\noindent sbrk($n$) ~~~~~~~~~~~~ \# Grow memory by $n$ bytes

\noindent -- -- -- -- -- -- -- -- -- -- -- -- -- -- -- -- -- -- -- -- -- -- -- -- -- -- -- 

$\mathbf{P} ~ {\oplus}{=} ~ \mathbb{A}(\mathbf{p},{\rm memory},n)$ ~ \# Add $n$ bytes of memory

\noindent --------------------------------------------------------------------

\noindent --------------------------------------------------------------------

\noindent chdir(dir) ~~~~~~~~~~ \# Change current directories

\noindent -- -- -- -- -- -- -- -- -- -- -- -- -- -- -- -- -- -- -- -- -- -- -- -- -- -- -- 

$\mathbf{P} ~ {\oplus}{=} ~ \mathbb{A}(\mathbf{p},{\rm cwd},{\rm dir})$ ~~~~~ \# Add new dir

\noindent --------------------------------------------------------------------

\noindent --------------------------------------------------------------------

\noindent $\dot{\mathbf{f}}$ = open($\dot{\mathbf{F}}$) ~~~~~~ \# Open files $\dot{\mathbf{F}}$

\noindent -- -- -- -- -- -- -- -- -- -- -- -- -- -- -- -- -- -- -- -- -- -- -- -- -- -- -- 

$\mathbf{P} ~ {\oplus}{=} ~ \dot{\mathbf{F}} \oplus \mathbb{A}(\dot{\mathbf{f}},{\rm open},1)$ ~ \# Mark as open

\noindent --------------------------------------------------------------------

\noindent --------------------------------------------------------------------

\noindent close($\dot{\mathbf{f}}$) ~~~~~~~~~~~ \# Close file IDs $\dot{\mathbf{f}}$

\noindent -- -- -- -- -- -- -- -- -- -- -- -- -- -- -- -- -- -- -- -- -- -- -- -- -- -- -- 

$\mathbf{P} ~ {\oplus}{=} ~ \text{-}\mathbb{A}(\dot{\mathbf{f}},{\rm open},1)$ ~~~~~~ \# Remove open flags

\noindent --------------------------------------------------------------------

\noindent --------------------------------------------------------------------

\noindent $\dot{\mathbf{F}}$ = fstat($\dot{\mathbf{f}}$) ~~~~~~ \# Get metadata on $\dot{\mathbf{f}}$

\noindent -- -- -- -- -- -- -- -- -- -- -- -- -- -- -- -- -- -- -- -- -- -- -- -- -- -- -- 

$\dot{\mathbf{F}} = \mathbb{I}(\dot{\mathbf{f}}) \mathbf{F}$ ~~~~~~~~~~~~~ \# Get corresponding rows

\noindent --------------------------------------------------------------------

\noindent --------------------------------------------------------------------

\noindent mkdir($\dot{\mathbf{F}}$) ~~~~~~~~~ \# Make diretories $\dot{\mathbf{F}}$

\noindent -- -- -- -- -- -- -- -- -- -- -- -- -- -- -- -- -- -- -- -- -- -- -- -- -- -- -- 

close(open($\dot{\mathbf{F}}$)) ~~~~~~~~ \# Open and close to create 

\noindent --------------------------------------------------------------------

\noindent --------------------------------------------------------------------

\noindent $\dot{\mathbf{F}}$ = dup($\dot{\mathbf{f}}$) ~~~~~~ \# Duplicate file descriptors

\noindent -- -- -- -- -- -- -- -- -- -- -- -- -- -- -- -- -- -- -- -- -- -- -- -- -- -- -- 

$\dot{\mathbf{f}} ~~ = $ \ldots ~~~~~~~~~ \# Create new file IDs

$\mathbf{F} ~ {\oplus}{=} ~ \mathbb{I}(\dot{\mathbf{f}},\mathbf{f}) \mathbf{F}$ ~ \# Copy file metadata to new IDs

\noindent --------------------------------------------------------------------

\noindent --------------------------------------------------------------------

\noindent link($\dot{\mathbf{F}}$) ~~~~~~~~~~ \# Create links to files

\noindent -- -- -- -- -- -- -- -- -- -- -- -- -- -- -- -- -- -- -- -- -- -- -- -- -- -- -- 

$\mathbf{F} ~ {\oplus}{=} ~ \dot{\mathbf{F}}$ ~~~~~~~~ \# Copy file metadata to new files

\noindent --------------------------------------------------------------------

\noindent --------------------------------------------------------------------

\noindent unlink($\dot{\mathbf{F}}$) ~~~~~~ \# Unlink files

\noindent -- -- -- -- -- -- -- -- -- -- -- -- -- -- -- -- -- -- -- -- -- -- -- -- -- -- -- 

$\mathbf{F} ~ {\oplus}{=} ~ \text{-}\dot{\mathbf{F}}$ ~~~~~~ \# Remove file metadata from files

\noindent --------------------------------------------------------------------

\noindent --------------------------------------------------------------------

\noindent mknod($\dot{\mathbf{F}}$) ~~~~~~~~ \# Make nodes to devices $\dot{\mathbf{F}}$

\noindent -- -- -- -- -- -- -- -- -- -- -- -- -- -- -- -- -- -- -- -- -- -- -- -- -- -- -- 

close(open($\dot{\mathbf{F}}$)) ~~~~~~~~ \# Open and close to create

\noindent --------------------------------------------------------------------

\noindent --------------------------------------------------------------------

\noindent pipe($\dot{\mathbf{F}}$) ~~~~~~~~ \# Make pipe $\dot{\mathbf{F}}$

\noindent -- -- -- -- -- -- -- -- -- -- -- -- -- -- -- -- -- -- -- -- -- -- -- -- -- -- -- 

close(open($\dot{\mathbf{F}}$)) ~~~~~~~~ \# Open and close to create 

\noindent --------------------------------------------------------------------

\end{document}